Table-top Measurement of Local Magnetization Dynamics Using Picosecond Thermal Gradients: Toward Nanoscale Magnetic Imaging

J. M. Bartell[1,†], D.H. Ngai[1,†], Z. Leng[1], and G. D. Fuchs[1]

[1]Cornell University, Ithaca, NY 14853

[†]These authors contributed equally to this work

**Recent advances in nanoscale magnetism have demonstrated the potential for spin-based technology including magnetic random access memory[1,2], nanoscale microwave sources[3,4], and ultra-low power signal transfer[5]. Future engineering advances and new scientific discoveries will be enabled by research tools capable of examining local magnetization dynamics at length and time scales fundamental to spatiotemporal variations in magnetic systems[6] – typically 10-200 nm[7,8] and 5 – 50 ps. A key problem is that current table-top magnetic microscopy cannot access both of these scales simultaneously. In this letter, we introduce a spatiotemporal magnetic microscopy that uses magneto-thermoelectric interactions to measure local magnetization via the time-resolved anomalous Nernst effect (TRANE). By generating a short-lived, local thermal gradient, the magnetic moment is transduced into an electrical signal. Experimentally, we show that TRANE microscopy has time resolution below 30 ps and spatial resolution limited by the thermal excitation area. Furthermore, we present numerical simulations to show that the thermal spot size sets the limits of the spatial resolution, even at 50 nm. The thermal effects used for TRANE microscopy have no fundamental limit on their spatial resolution, therefore a future TRANE microscope employing a scanning plasmon antenna could enable measurements of nanoscale magnetic dynamics.**

Magnetic microscopy has played a fundamental role in the study of magnetic behavior such as domain wall motion[9,10], skyrmion formation[11], magnetic switching[12], and spin wave propagation[13], spurring interest in the dynamics of nanoscale magnetic features[14,15]. This motivates a method of spatiotemporal magnetic microscopy capable of measuring picosecond changes in the magnetic moment with spatial resolution of less than 200 nm[7,8]. Magneto-optical measurements are currently the only table-top approach to measure spatially varying dynamics in the time-domain. Unfortunately, the spatial resolution available to optical measurements is limited by diffraction to approximately $\lambda/(2\,NA)$, where $\lambda$ is the wavelength of light and $NA$ is the effective numerical aperture of the focusing optics. Therefore, optical techniques including the time-resolved magneto-optical Kerr effect (TRMOKE), have a diffraction limited resolution of roughly 200 nm. One solution is to use nanometer-scale wavelength radiation, as in X-ray magnetic circular dichroism (XMCD) experiments which provide spatial resolution of 30 nm and time-domain resolution of less than 100 ps[16]. Unfortunately, spatiotemporal XMCD requires synchrotron-based sources which limits its wide-spread use.

To circumvent the spatial limitation imposed by optical diffraction, we propose a new technique for magnetic spatiotemporal microscopy that is based on the interaction between magnetization and heat rather than light. Our method is based on the time-resolved anomalous Nernst effect (TRANE). The geometry for TRANE is depicted in Fig. 1a. The anomalous Nernst effect (ANE) is a magnetization dependent, thermoelectric effect[17–19], in which a thermal gradient, transverse to the film's magnetic moment, generates an electric field given by[20], $\vec{E}_{ANE}(\vec{x},t) = -N\,\vec{\nabla}T(\vec{x},t) \times \mu_o \vec{M}(\vec{x},t)$, where N is the anomalous Nernst coefficient. Previous studies have demonstrated that by confining $\vec{\nabla}T(\vec{x},t)$ to a micron-scale region in a thin-film ferromagnetic metal, an anomalous Nernst voltage is generated that is proportional to the local magnetic moment[21,22]. This has inspired proposals for applications that include microscopy and spectroscopy[21–24], yet, a fully developed spatiotemporal microscope with nanoscale resolution has not yet been realized. Here, we show that using a pulsed thermal source with a short duty cycle from a focused laser, $E_{ANE}$ can be localized in both time and space to generate a TRANE signal.

TRANE microscopy is a viable strategy for high spatiotemporal resolution because the ANE interaction time and the electron thermal carrier wavelength are both short in comparison to the scales of magnetic dynamics and the spatial variation of magnetization. Because thermal gradients are not fundamentally limited by optical diffraction, microscopy based on magneto-thermal interactions has no fundamental barrier to decreasing the spatial resolution. Therefore, the spatiotemporal resolution of TRANE microscopy is predominantly limited by the generation and evolution of the localized thermal gradient.

Figure 1b shows a schematic for the measurement setup. We focus a pulsed laser to generate the short-lived, local temperature gradient for each optical pulse, thus creating a corresponding voltage pulse. Using a homodyne technique by electrically mixing the generated voltage pulse and lock-in measurement, we measure the TRANE signal, $V_{TRANE}$, which is proportional to the stroboscopically sampled local magnetization. In this work, we used optical pulses with a fluence of 2.3 mJ/cm$^2$ that created vertical thermal gradients of $3.3 \times 10^8$ K/m and a temperature increase of 30 K at the hottest point (see supporting online information)[25]. The first structure we use to demonstrate TRANE microscopy is a 30 nm cobalt film patterned in 18 μm wide cross-structures. Fig. 1c shows a hysteresis curve of this sample measured using TRANE, which demonstrates the proportionality between $V_{TRANE}$ and the local magnetic moment.

To show that the optically generated thermal gradients are short-lived, we time-resolved $V_{TRANE}$ by mixing it with a short, 75 ps electrical probe pulse. Fig. 1d shows our measurement of $V_{TRANE}$ as a function of electrical pulse delay. These data can be understood as the temporal convolution of $V_{TRANE}$ with the electrical probe pulse[25]. We find that the convolution also has a 75 ps width, which we interpret as the upper limit of the thermal gradient's lifetime. As we show with subsequent magnetic resonance experiments, thermal gradients produced in our microscope are actually much shorter-lived.

The sensitivity of TRANE microscopy is dependent on several factors, including the Nernst coefficient, the geometry and the impedance. For the 18 µm cross structure, we calculate the magnetization angle sensitivity to be $0.73°/\sqrt{Hz}$ [25]. The electrical TRANE signal scales as $d^2/w$ for a probe diameter, $d$, and a channel width, $w$[21]. For comparison, the signal scaling of magneto-optical microscopy is essentially independent of the device geometry above the optical diffraction limit of $d \sim \lambda/(2\ NA) \sim 200$ nm. It is below this fundamental limit that resolution of far-field magneto-optical microscopy is sharply reduced. In contrast, a TRANE signal collected with a nanoscale probe diameter (<200 nm) can remain large provided that $w$ is also scaled. Because of its picosecond duration, the TRANE signal is also sensitive to the microwave impedance of the sample, with the strongest signal occurring when the sample impedance matches the 50 Ω impedance of the measurement circuit.

Next we experimentally demonstrate that lateral thermal diffusion does not limit spatial resolution at the scale of a tightly focused laser by imaging the local magnetic moment of the cobalt cross. Scanning the laser across the sample, a map of the magnetization is created (shown in Fig. 2a) in which, domain walls are visible where the projected moment is zero. For the cobalt films studied here, the domain walls are 150-200 nm wide[26], which is far below the 440 nm Abbe resolution limit we calculate for our apparatus. We use this fact to evaluate the resolution of our TRANE microscope by fitting spatial line cuts across a magnetic domain wall with a convolution of a step function and a Gaussian function of width, 2δ. The fit yields δ =460 ± 90 nm[25]. These results suggest that the main limitation to the spatial resolution is the size of the thermal gradient spot.

To gain a deeper understanding of thermal diffusion in our magnetic thin film samples, we performed time-dependent, finite element simulations of the picosecond heating dynamics. Using numerical simulations, we show in Fig. 2d that when the laser pulse is at its maximum, the gradient's vertical component of the thermal gradient does not spread laterally beyond the pulsed heat source. Interestingly, our simulations show this statement is true even for nanoscale diameter thermal sources. Although the thermal spot size used in this demonstration is limited by optical diffraction, the thermal gradient diameter could be reduced below the far-field optical diffraction limit using a light-confining plasmon antenna[27,28].

We study TRANE's temporal resolution by stroboscopically measuring ferromagnetic resonance (FMR) in $Ni_{20}Fe_{80}$ (permalloy) wires using the apparatus depicted in Fig. 3a. The wire axis of the sample is aligned parallel to the applied magnetic field. To excite magnetization dynamics, a microwave frequency current is passed through a nearby copper wire to generate an out-of-plane magnetic field on the permalloy. We choose a drive frequency that is commensurate with the laser repetition rate to create a constant phase relationship between the $V_{TRANE}$ measurement of magnetization and the excitation field. Starting with the detection scheme as before, we also modulate the external magnetic field to distinguish $V_{TRANE}$ from other voltages due to inductive coupling between the two wires (see methods).

In Fig. 3b we plot FMR as a function of magnetic field. The magnetization is excited by a 5.00 GHz stimulation to a maximum oscillation angle of angle of 0.07°. The measurement sensitivity is $0.093°/\sqrt{Hz}$, which is improved from the cobalt cross sample chiefly because of the reduced sample width that increases the ratio $d^2/w$. By electrically shifting the relative time delay between the microwave magnetic field drive and the laser probe by 50 ps, these data also demonstrate that TRANE microscopy is sensitive to the phase of magnetic precession.

The FMR data are analyzed by fitting to linear combinations of symmetric and anti-symmetric Lorentzian functions modified to account for the magnetic field modulation (see SI). From the fits, we extract a phase difference between the two of 64° ± 24°. The discrepancy from our expectation of a 90° shift might be because our simple model accounts only for a single, uniform FMR mode. Close inspection of the two data sets in Fig. 3b reveals additional features that are anti-correlated between measurement phases. This suggests more complicated magnetic behavior than we model, including the existence of additional magnetic modes that may influence the accuracy of the phase we extract from fitting. Although full imaging and analysis of these modes is a capability of TRANE microscopy, their detailed study is beyond this scope of the present demonstration.

As we increase the frequency of the magnetic excitation, we find that (as expected) the FMR resonance field evolves as described by the Kittel equation, $\frac{\gamma}{2\pi}\sqrt{(H + N_z M_s)(H + N_y M_s)}$, as shown in Fig. 3d. Here, we use $N_y = 0.015$, and $N_z = 0.985$, which are determined separately with measurements of the hard axis magnetic saturation. From these fits we find an effective magnetic moment $4\pi M_s$ = 840 emu/cm$^3$ and a Gilbert damping parameter, $\alpha = 0.009 \pm 0.001$. The damping in this sample is consistent with separate FMR measurements that we made by electrically monitoring the DC rectification voltage. These results are also in excellent agreement with literature values for permalloy[29,30]. The consistency among our various measurements and prior reports supports the proposal that the local, transient heating of the sample during measurement does not significantly alter its dynamical properties as probed by TRANE microscopy.

To experimentally determine the time scale of the vertical thermal gradient decay, we measure FMR as we increase the stimulation frequency. Assuming that the thermal pulses sample the mean magnetic projection over their duration, observation of FMR at a particular frequency sets an upper limit on the thermal decay time. In the inset to Fig. 3c we plot the FMR spectra at 16.4 GHz, which is the highest frequency that we can produce with our microwave electronics. From these data, we conclude that the thermal gradient decays in under 30 ps. This is supported by our time-dependent finite element modeling (Fig. 3c) which shows the thermal gradient pulse has a full width at half-maximum of 10 ps for these samples. Therefore, this technique could potentially be extended to measure FMR dynamics up to 50 GHz.

In conclusion, we demonstrated that TRANE microscopy enables a pathway to resolving magnetic dynamics on fundamental length and time scales. We have demonstrated that for the thin film samples studied here, TRANE microscopy has temporal resolution below 30 ps and spatial resolution at the limit of focused light. As a magneto-thermoelectric technique, TRANE microscopy is not subject to the fundamental diffraction limit of spatial resolution that constrains far-field optical methods. Crucially, the numerical simulations indicate that the spatial resolution is only determined by the lateral diameter of the thermal gradient, which we have verified down to a diameter of 50 nm. Applying these capabilities in a table-top imaging platform can enable new access to time-resolved magnetization dynamics which supports the burgeoning field of high-speed magnetoelectronics.

**Methods**

*Sample Preparation*

For measurements of spatial resolution, 30 nm thick cobalt films were deposited by electron beam evaporation onto sapphire substrates. Photolithography and ion milling was used to pattern the films into square crosses as pictured. For the spatial images presented, the cross arms were 18 μm wide. Electrical contact was made by wire bonding to evaporated copper contacts.

The samples used for ferromagnetic resonance measurements were 30 nm thick $Ni_{20}Fe_{80}$ (permalloy) films deposited by DC magnetron sputtering at a base pressure below $10^{-7}$ Torr. The films were patterned with e-beam lithography and ion milled into wires 2 μm wide and 950 μm long. Evaporated copper contacts 1 μm wide were fabricated to contact the permalloy wire with a range of separations to enable a DC impedance match close to 50 Ω. The contacts chosen for the measurement were 3 μm apart and had a DC resistance of 74 Ω. The wire we used as a microwave antenna to excite magnetization dynamics was fabricated in a lift-off process to be 2 μm wide, 50 μm long, and 102 nm thick. It was positioned 1 μm away from the permalloy wire and had a DC resistance of 48 Ω.

*Thermal Gradient Generation*

Local thermal gradients were generated by focusing light from a Titanium:Sapphire laser tuned to 794 nm with 3 ps pulses and a fluence at the sample of 2.3 mJ/cm$^2$. The repetition rate was controlled with an electro-optic modulator/pulse picker. We used a repetition rate of 76 MHz for measurements of the spatial imaging and 25.33 MHz for the ferromagnetic resonance measurements. An optical chopper was used to modulate the optical pulse train at 9.7 kHz. To scan the beam, we used a 4-F optical path in combination with a voice-coil controlled fast-steering mirror. The light was focused into a diffraction-limited spot using a 0.90 numerical aperture air objective.

*Detection*

To detect the TRANE voltage pulses, we connect the voltage contact to a microwave transmission line through a coplanar waveguide soldered to a type-K connector. The signal is passed through a low-pass filter with a 4 GHz break frequency to attenuate GHz frequency artifacts from inductive electrical coupling between the copper antenna and permalloy wire. After the filter, the signal is amplified by 40 dB with a 0-1 GHz bandwidth. The amplified pulse train is sent to the RF port of an electrical mixer, where it is mixed with a 1.5 ns pulse train from a pulse/pattern generator that is referenced to the laser repetition rate. When the two pulse trains temporally overlap, a voltage modulated by the optical chopper (and, for FMR, the field modulation) is passed to a low-frequency preamplifier before being sent to a lock-in amplifier.

To determine the amplitude of $V_{TRANE}$ prior to amplification and electrical mixing, we calibrated the transfer function of the "collection" circuit by measuring the signal produced by electrically generated reference pulses and systematically varying their widths. We find that our collection circuit transfer coefficient is $0.47 \pm 0.04$ for a 10 ps signal pulse (see SI). Using this calibration, we measure that the anomalous Nernst coefficient in permalloy is $2.4 \pm 0.2 \times 10^{-6}$ VT$^{-1}$K$^{-1}$, which agrees with a previous report[20].

*2D Imaging*

Imaging the static magnetic moment is performed by measuring the $V_{TRANE}$ along a channel perpendicular to the applied magnetic field so that the maximum signal was obtained during saturation of the magnetic moment. The multi-domain state was prepared by saturating the cross with a 130 Oe field and decreasing the field to 32 Oe. For the data in Fig. 2a, we used a 250 nm step and a lock-in time constant of 500 ms.

*FMR Excitation*

FMR was excited in the samples using a microwave signal produced by an arbitrary waveform generator (AWG) with a clock referenced to the laser repetition rate. This clock is multiplied up within the AWG to a sampling rate of 19.98 GS/s derived from the 25.3 MHz laser pulse repetition rate. The waveforms from the AWG can be delayed in steps of 50 ps with respect to the laser pulses without re-triggering, allowing resonant behavior of different phases to be observed. For excitation frequencies above 5.7 GHz, the output frequency of the AWG was doubled or quadrupled with electrical frequency multipliers to achieve frequencies up to 16.4 GHz. This excitation signal was then amplified to a power between 13-20 dBm and coupled to the copper stimulation wire.

*TRANE detection of FMR*

Ferromagnetic resonance was detected by using a second lock-in with dual demodulation. In this technique, two modulation sources at different frequencies are used. The signal is extracted by first demodulating the input referenced to the optical chopper. The resulting signal is then sent to a second demodulator (time constant of 1 s) that is referenced to a 5-10 Hz modulation of the magnetic field.

**Acknowledgements**

This work was supported by AFOSR. The authors would like to thank Isaiah Gray for his help measuring the temperature dependence of permalloy. This work made use of the Cornell Center for Materials Research Shared Facilities which are supported through the NSF MRSEC program (DMR-1120296) as well as the Cornell NanoScale Facility, a member of the National Nanotechnology Infrastructure Network, supported by the NSF (Grant ECCS-0335765).

**Author Contributions**

GDF, DHN, and JMB developed the concept and procedure for the experiment. JMB assembled the optical apparatus, DHN fabricated the samples. ZL and JMB performed numerical simulations. JMB and DHN performed the experimental measurements. GDF, JMB, and DHN analyzed the data and wrote the manuscript.

**Figures**

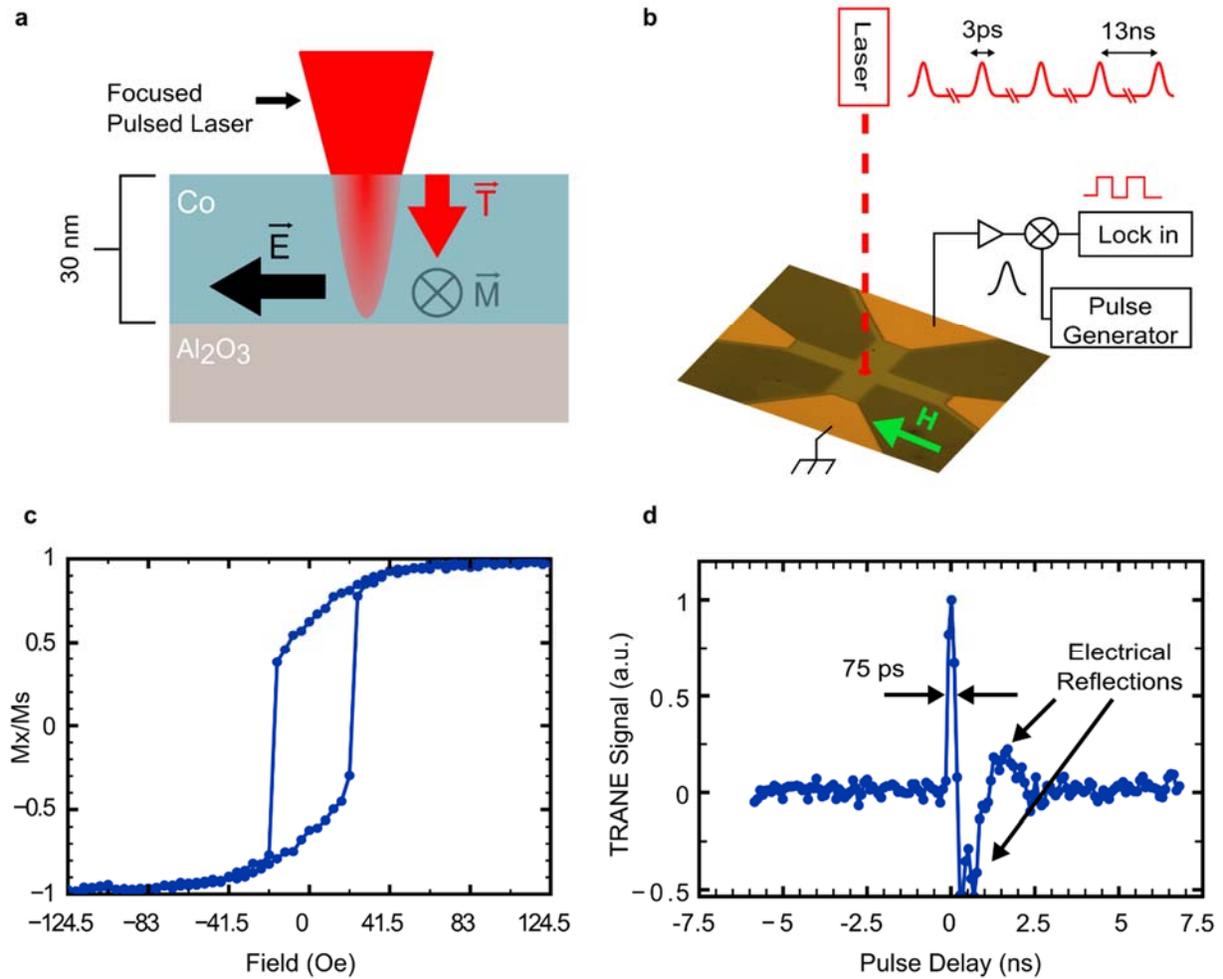

**Figure 1| The Time Resolve Anomalous Nernst Effect for Magnetic Measurement. a,** Schematic diagram of the anomalous Nernst effect in our measurements **b**, Schematic of the experimental setup. 792 nm pulsed laser light is focused to a diffraction limited spot on the magnetic film patterned on top of a thermally conductive, electrically insulating substrate. Bonding pads enable detection of the TRANE voltage proportional to the perpendicular magnetic moment. **c,** TRANE measurement of local hysteresis in the cobalt cross. **d,** Signal from mixing the TRANE voltage with a 75 ps electrical pulse. We observe that the width of the initial peak is

Page 16

75 ps, indicating that the TRANE pulse is < 75 ps. The other, broader features are electrical reflections due to the impedance mismatch of the device and the 50 Ω transmission line.

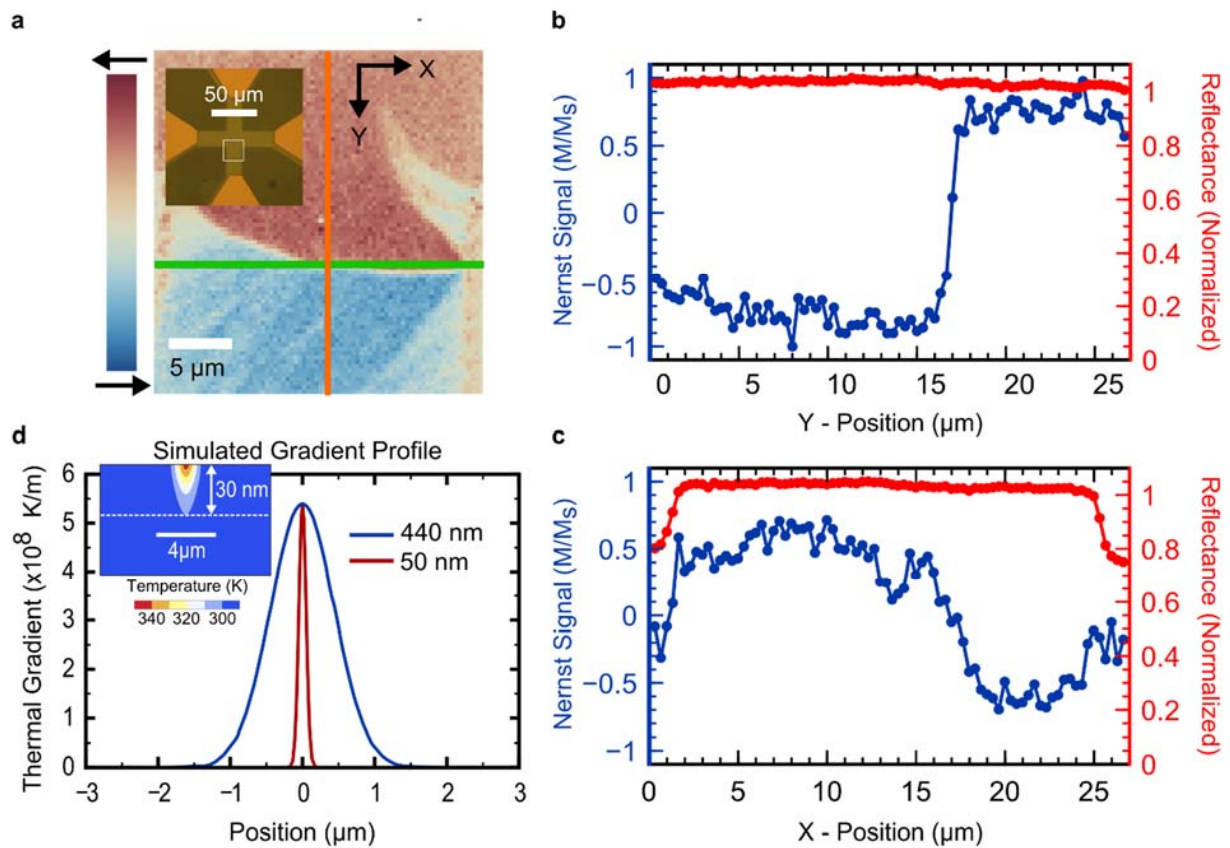

**Figure 2 | Spatial Resolution of Magnetic Imaging Using a Thermal Gradient. a,** Image of magnetic structure taken with TRANE. **b & c,** Line cuts of the image showing the TRANE signal in blue and the simultaneously measured reflected light in red. Top: line cut from top to bottom. Bottom: line cut from left to right we note that the reflected signal drops off at the edge of the cross while the TRANE signal goes to zero without edge artifacts. **d,** Finite element, time dependent simulation of the vertical component of the thermal gradient at peak applied power along the x-axis as a function of distance from the spot center. The blue and red temperature profiles correspond to a thermal generation spot size of 440 nm and 50 nm respectively. The inset shows the x-z profile of the temperature at the center of the optical pulse. We note that, because of radial symmetry, the radial (in-plane) gradient gives no signal.

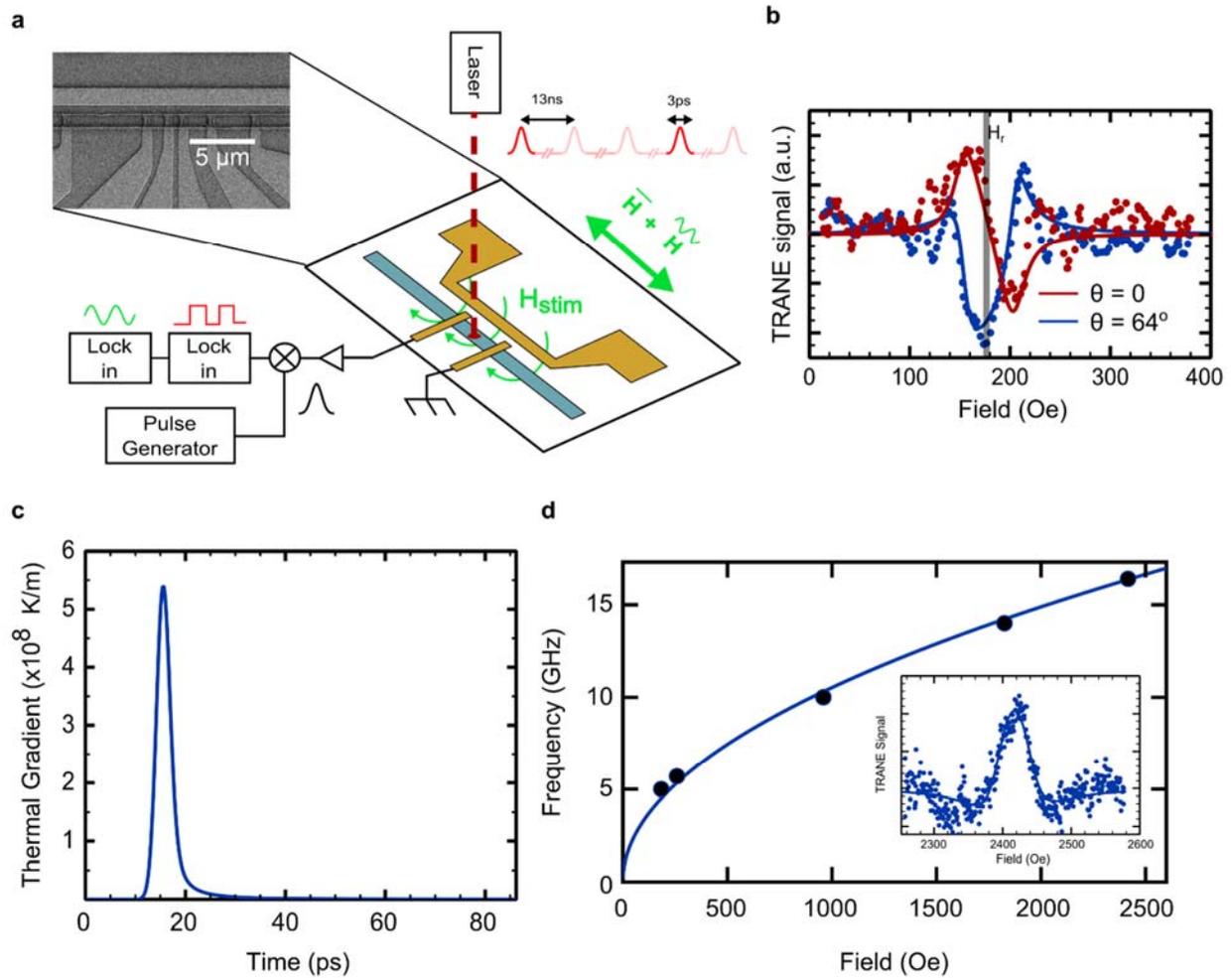

**Figure 3 | Measurement of Magnetic Dynamics using TRANE. a,** Schematic of the experimental setup used to measure FMR in permalloy wires. The 2 μm wide permalloy wire (blue) is stimulated by a 2 μm Cu wire 1 μm away from the permalloy wire. The contacts used to measure the TRANE signal were separated by 3 μm. **b,** FMR spectra of taken at 5 GHz and two different phases of the stimulation. The solid lines are fits to the Lorentzian curves after accounting for the modulated lock-in technique. **c,** Time dependent numerical simulation of the thermal gradient as a function of time. **d,** Plot of the resonant frequencies as a function of the resonant field determined by fitting. The solid line is a fit to the Kittel equation. FMR spectra

taken at 16.4 GHz is shown in the inset. For the FMR spectra, the points show the data after smoothing over 3 neighboring points. The solid line is a fit to the linear combination of symmetric and anti-symmetric Lorentzians after accounting for the modulation frequency.

**Supplemental Information**

**1. Determination of Laser Induced Temperature Change**

The increase in local sample temperature due to laser heating can be measured from the resulting increase in local resistance. Using the same detection scheme as depicted in Fig. 1b, we measure the voltage change due to the heating as a function of an applied DC current. The voltage measured is given by $V = \beta\, I_{DC}\, \Delta R$, where $\beta$ is the collection circuit transfer coefficient, $I_{DC}$ is the applied DC current and $\Delta R$ is the resistance change due to laser heating. Note that as discussed in a later section, $\beta$ varies for the pulse width. Thus we have different transfer coefficients for the overall temperature and the temperature gradient due to the different decay times. By relating the resistance change to a temperature increase, we can quantitatively determine the heating induced by the laser.

To relate the resistance change to the temperature profile, we first measure the temperature dependence of the 4-terminal sample resistivity. For our permalloy sample in the temperature range of interest, the resistivity is given by a linear temperature dependence given by $\rho(T) = \rho_0(1 + \alpha\,(T - T_0))$ with $\alpha = 0.0025\ K^{-1}$, $\rho_0 = 30\ \Omega\ cm^{-2}$ and $T_0 = 293\ K$. Using the temperature profile calculated from numerical simulations described in the following section, we measure the maximum temperature increase at various fluences as shown in Fig. S1. For the typical fluence of 2.3 mJ/cm² used throughout the measurement, we have a maximum temperature increase of 30 K. Using the simulated temperature profile and maximum signal from the hard-axis hysteresis, we measure an anomalous Nernst coefficient of $2.4 \pm 0.2 \times 10^{-6}$ V/(T K). This measurement is in agreement with a recently reported value[1].

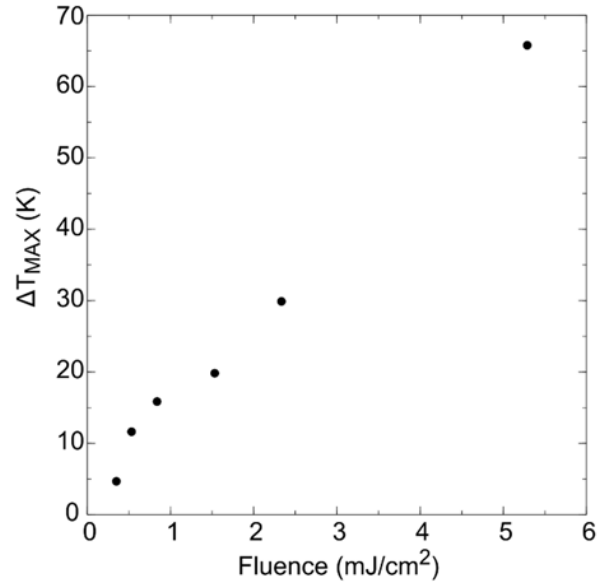

**Figure S1 | Maximum Temperature Increase due to the Laser.** The experimentally measured temperature increase as a function of laser fluence. For TRANE measurements, we used a fluence of 2.3 mJ/cm$^2$, which corresponds to 30 K increase in temperature.

## 2. Finite Element Simulations of Thermal Evolution

Finite element modeling of the thermal gradient evolution was performed using the COMSOL Multiphysics Heat Transfer Module. We consider a single temperature diffusive model in which the laser is treated only as a heat source, rather than considering different the phonon and electrons temperatures. This is justified by the fact that the optically excited electrons are thermalized on time scales comparable to the laser pulse width of 3 ps[2].

The spatiotemporal evolution of the thermal gradient in our system is calculated numerically with the Fourier diffusion equation using the material parameters given in Table S1. The heat source $Q(\vec{x}, t)$, is given by,

$$Q(\vec{x},t) = \frac{Q_o}{2\pi\,\delta_x\,\delta_y\,d}\, e^{-\frac{x^2}{2\delta_x^2}-\frac{y^2}{2\delta_y^2}}\, e^{-\frac{z}{d}} e^{-\frac{t^2}{2\tau^2}} \tag{S1}$$

where, $\delta_x$ and $\delta_y$ are the Gaussian widths in the x and y direction of the laser spot (440 nm), d is the skin depth (12nm), $Q_o$ is the incident peak power of a single pulse (2.19 W), and $\tau$ is the pulse Gaussian temporal width of the 3 ps pulse.

**Table S1 | Material Parameters used for Simulation**

| Material | Thermal Conductivity (W/m K) | Specific Heat (J/g K) | Density (g/cm$^3$) |
|---|---|---|---|
| Sapphire[3] | 30.3 | 0.764 | 3.98 |
| Permalloy | 46.4[4] | 0.43[5] | 8.7[6] |

The results of the simulations yield spatiotemporal profiles of the temperature and thermal gradient shown in Fig. S2. To apply the simulation for quantitative analysis we need a sample specific scaling factor determined experimentally (See supplemental sections 1 and 6). For the permalloy samples presented in this letter the scaling factor was found to be 0.47 ± 0.04.

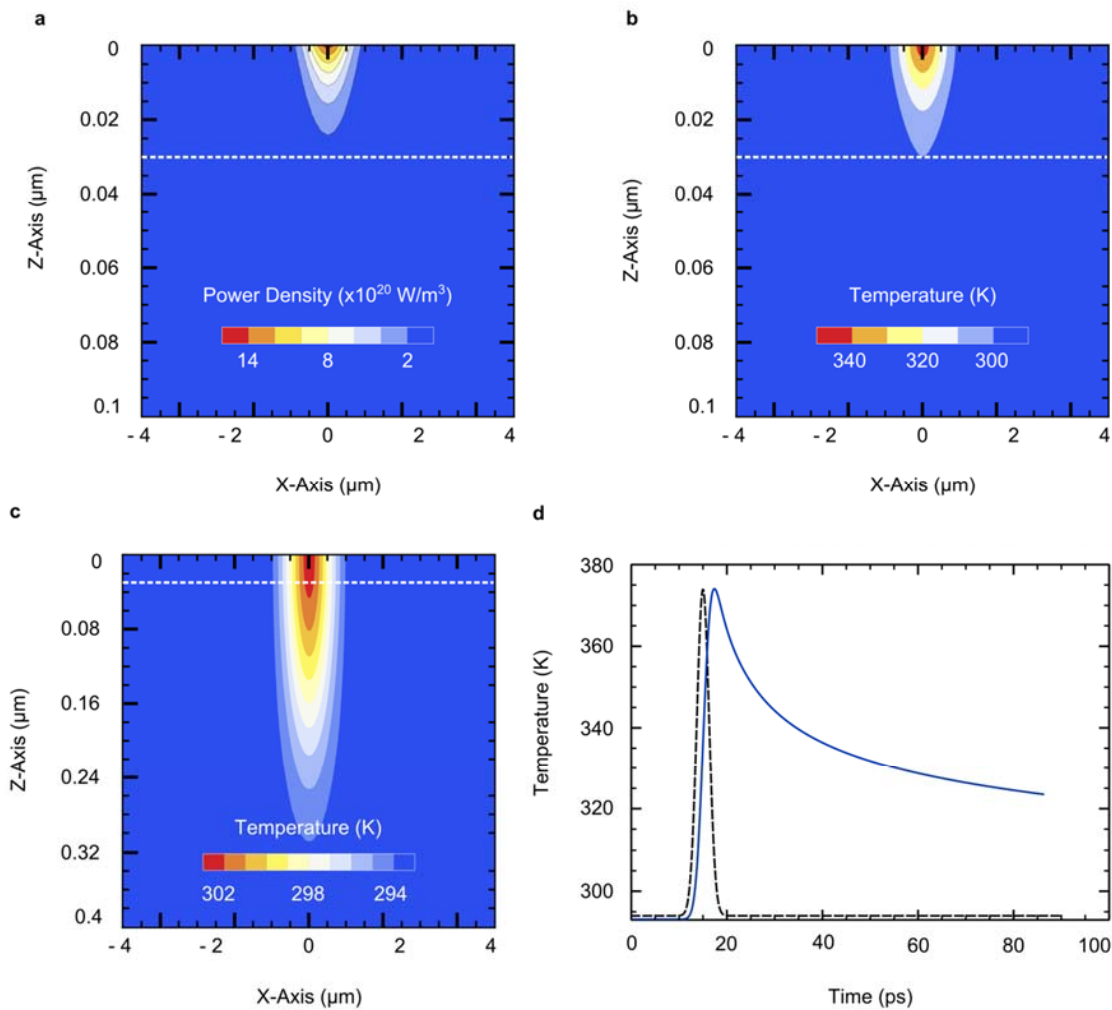

**Figure S2 | Simulated Spatial and Temporal Temperature Profiles. a,** Spatial profile of the heat source, $Q(\vec{x}, t)$, for the 440 nm spot size. **b-c,** Temperature profiles across an axial slice of the thermal source of the 440 nm Gaussian width. The dashed line indicates the interface between the permalloy wire and the sapphire. **b,** is the temperature at the peak of the pulse and **c,** is the temperature 982 ps after the peak. **d,** Time dependence of the laser induced temperature increase for 440 nm spot size. The dashed line shows the temporal profile of the heat source in arbitrary y – axis for reference.

Page 24

## 3. Fitting Lateral Resolution

We measure the value for the lateral resolution by taking vertical line cuts of the 2D scan across a portion of the domain wall (Fig. 2a in the main text and Fig. S3a). The 4 μm region of the domain wall used for fitting is shown boxed in Fig. S3a. This region was chosen because it was the portion of the image with the clearest step function behavior. Fits of the line scans where done using a least means squared method to find the Gaussian width, amplitude, and center of a function derived by convolution of a Gaussian with a –1 to 1 step function. The results of the individual fits are shown in Fig. S3b. The mean of the fits is 460 nm with a standard deviation of 90 nm, the standard deviation is used as the uncertainty, as it was larger than the uncertainty of the individual fits.

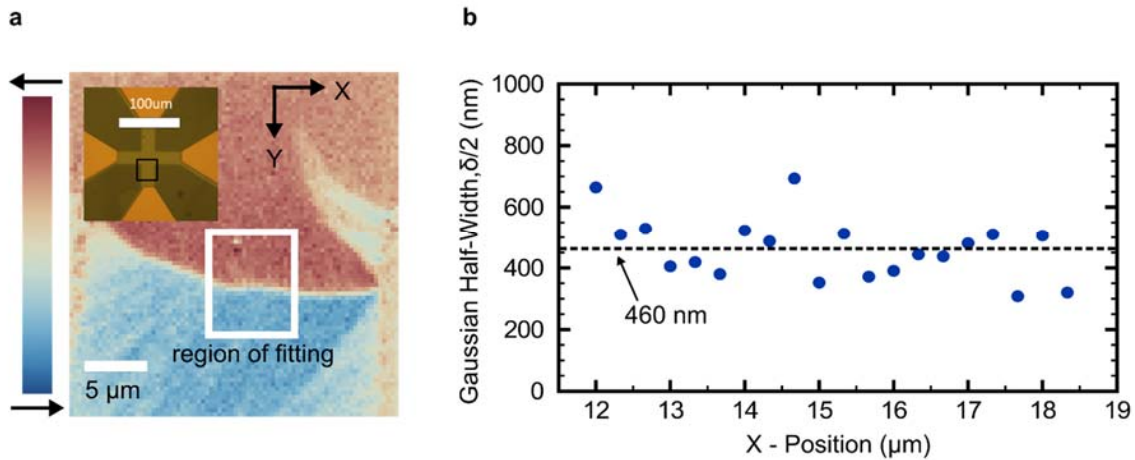

**Figure S3 | Region and Results of Spatial Fitting. a**, Spatial map of the static magnetic moment showing the region used for the lateral resolution measurement. **b,** One-half the Gaussian-width of the pulse that was convolved with the unit step determined by fitting. The X-axis is the horizontal coordinate of the line cut used and the dashed line indicates the mean.

## 4. Modification of the Resonant Line-shapes due to Field Modulation

To measure the FMR of the permalloy wires we detect the projected magnetic moment perpendicular to the wire. The magnetic moment of the wire precesses about the externally applied magnetic field when driven by an external microwave field from a microwave antenna patterned parallel to the magnetic wire. The FMR precession angle of a ferromagnetic in the linear response regime can be modeled as a driven damped oscillator. The projection amplitude of this motion is the linear combination of even and odd Lorentzian functions.

$$Sin(\varphi)\frac{(H-H_r)/\vartheta}{1+(H-H_r)^2/\vartheta^2} + Cos(\varphi)\frac{1}{1+(H-H_r)^2/\vartheta^2} \tag{S2}$$

In addition to the magnetic signal due to FMR, we also detect an induced electrical response from coupling between the microwave antenna and the magnetic bar. To separate the two signals, a cascaded lock-in technique was used in which the first demodulation was referenced to a square modulated 9.7 kHz signal from an optical chopper and the second demodulation was referenced to a 14 Oe sinusoidal field modulated at 10 Hz (5 Hz for FMR frequencies above 10 GHz). The TRANE signal detected by the second lock-in can be modeled by Eq. S3.

$$Sin(\varphi)\int\frac{(H+H_{mod}Cos(\omega\,t)-H_r)/\vartheta}{1+(H+H_{mod}Cos(\omega\,t)-H_r)^2/\vartheta^2}*Cos(\omega\,t)dt$$
$$+ Cos(\varphi)\int\frac{1}{1+(H+H_{mod}Cos(\omega\,t)-H_r)^2/\vartheta^2}*Cos(\omega\,t)\,dt \tag{S3}$$

The resulting analytical equation is then used to fit the resonance data obtained with TRANE to quantify the values of the linewidth, amplitude, phase, and center frequency. We note that the modification to the Lorentzian shape does not add free parameters to the fitting function because the modulation amplitude is a known value. The modulation does impact the uncertainty and it reduces the overall signal amplitude, but at the benefit of increased angular sensitivity.

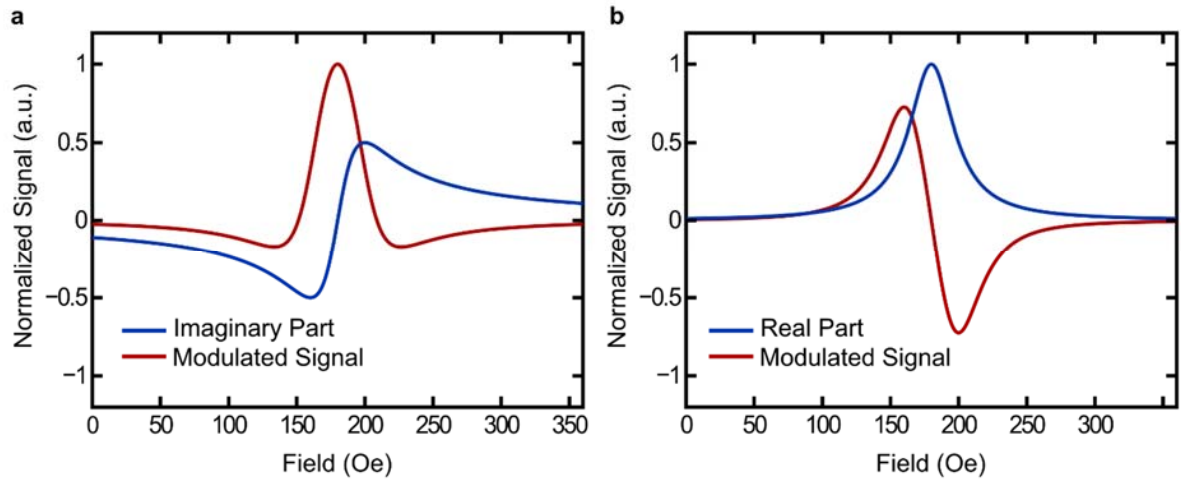

**Figure S4 | Modification of Lorentzian Functions.** The blue curve in each plot shows the modeled normalized - Lorentzian response function for the projected amplitude of the FMR precession for a resonant field, $H_r$ = 180 Oe and line-width of 80 Oe. The red curves show the corresponding signal as detected by the lock-in when using a modulation amplitude of 20 Oe.

## 5. Sensitivity

The sensitivity is calculated using the field-dependent magnetization measurements shown in Fig. 1c in the main text and Fig. S5. This measurement is done in the transverse geometry – the saturated moment is perpendicular to the voltage pick-ups – so that ($V_{TRANE}^{max} - V_{TRANE}^{min}$) corresponds to a 180° rotation. The standard deviation of points at saturation is taken as the detected voltage uncertainty, $\delta_{TRANE}$. As a longer sampling time will reduce the value of $\delta_{TRANE}$ regardless of the sample, it is desirable to have a sensitivity figure of merit independent of the sampling time. Thus, the signal-to-noise ratio by the measurement rate, in the case of a lock-in measurement this is the time constant. This yields an equation for the minimum detectable angle, $\theta_{min}$, with respect to the angle of highest sensitivity, $\theta_o = 90°$, measurable with the TRANE technique.

$$\theta_{min} = \frac{\delta_{TRANE}}{Sin(\theta_o)(V_{TRANE}^{max} - V_{TRANE}^{min})/2} \sqrt{TC} \qquad (S4)$$

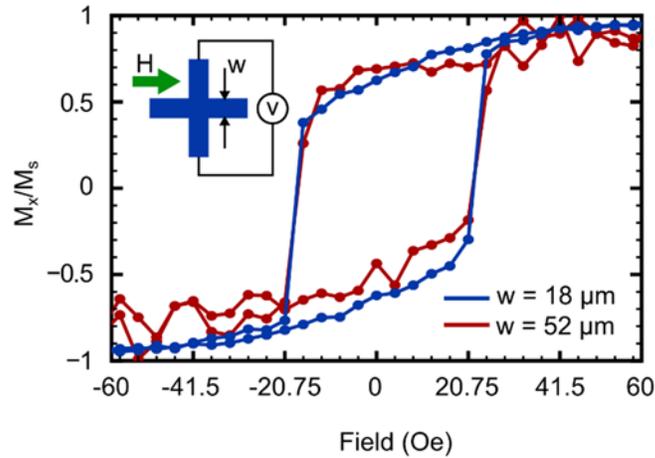

**Figure S5 | Comparison of Hysteresis Measurement Sensitivity.** In this graph we plot TRANE – measured hysteresis loops for two different cross sizes. We observe that the sensitivity is $\theta_{min} = 4.6°/\sqrt{Hz}$ for the cross with the 52 μm arms and $\theta_{min} = 0.73°/\sqrt{Hz}$ for the 18 μm wide cross.

## 6. Collection Circuit Transfer Coefficient

To determine the transfer coefficient of the collection circuit depicted in Fig. 1b and Fig. 3a in the main text, we measure the collection voltage from a calibration pulse. Numerical simulations suggests the voltage pulse from TRANE has a width of 10 ps. With the electronics available, we cannot create a 10 ps pulse to directly measure the transfer coefficient. Instead, we extrapolate it through measuring the gain of square pulses of wider widths. Fig. S6 shows the total gain in the collection circuit as a function of the square pulse width and the fit with our model.

To model the gain, we calculate the collection voltage from the amplified calibration square wave and pulse pattern generator. We treat the amplified calibration voltage into the mixer as a square wave with only its spectral components below 1 GHz due to the bandwidth limitations of the amplifier. The pulse pattern generator voltage is modeled as a triangle function with a width of 1.5 ns. The measured voltage after the mixer is the DC component of the multiplication of these two voltages. We fit this model to the data in Fig. S6 with the amplitude as the only free parameter. Extrapolating the fit, our model estimates a transfer coefficient of $0.47 \pm 0.04$ for a 10 ps pulse.

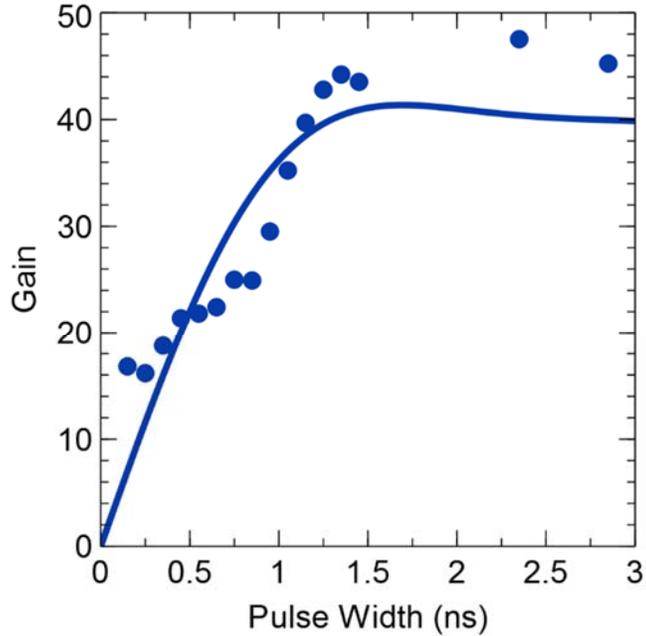

**Figure S6 | Collection Circuit Gain.** We plot the dependence of the collection circuit gain on the width of a calibrating square pulse. The model used to fit the data estimates a transfer coefficient of 0.47 ± 0.04 for a 10 ps TRANE pulse.

**Supplementary References**